\documentclass[12pt,english,floatfix,nofootinbib,superscriptaddress,aps,prd,preprint]{revtex4}
\usepackage[english]{babel}
\usepackage[utf8]{inputenc}
\usepackage{amsfonts}
\usepackage{amsmath}
\usepackage{amssymb}
\usepackage{graphicx,color}
\usepackage[bookmarks=true,colorlinks=true,linkcolor=blue,citecolor=blue,pdfstartview=FitV,urlcolor=blue,breaklinks=true]{hyperref}
\usepackage{indentfirst}
\usepackage{wrapfig}
\usepackage{float}
\usepackage{units}
\usepackage{textcomp}
\usepackage{wasysym}
\usepackage{multirow}
\usepackage{subfig}
\usepackage{array}
\usepackage{epsfig}
\usepackage{cancel}
\usepackage{stmaryrd}
\usepackage{upgreek}
\usepackage{subfig}
\usepackage{tabularx}
\usepackage{psfrag}
\usepackage{epsfig}
\usepackage{titletoc}
\usepackage{mathrsfs}
\usepackage[dvips]{feynmp}
\usepackage{fancyhdr}
\usepackage[dvipsnames]{xcolor}

\usepackage{hyperref}
\hypersetup{
    colorlinks,
    citecolor=blue,
    filecolor=green,
    linkcolor=purple,
    urlcolor=red,
}

\newcommand{\ie}{\begin{equation}}
\newcommand{\fe}{\end{equation}}
\newcommand{\se}{\begin{eqnarray}}
\newcommand{\ff}{\end{eqnarray}}

\begin{document}

\title{How does geometry affect quantum gases?}

\author{A. A. Ara\'{u}jo Filho}
\email{dilto@fisica.ufc.br}

\affiliation{Universidade Federal do Cear\'a (UFC), Departamento de F\'isica,\\ Campus do Pici,
Fortaleza -- CE, C.P. 6030, 60455-760 - Brazil.}

\author{J. A. A. S. Reis}
\email{joao.andrade@discente.ufma.br}

\affiliation{Universidade Federal do Maranh\~{a}o (UFMA), Departamento de F\'{\i}sica, Campus Universit\'{a}rio do Bacanga, S\~{a}o Lu\'{\i}s -- MA, 65080-805, Brazil}

\affiliation{Universidade Estadual do Maranh\~{a}o (UEMA), Departamento de F\'{i}sica,\\ Cidade Universit\'{a}ria Paulo VI, S\~{a}o Lu\'{i}s -- MA, 65055-310, Brazil}

\date{\today}

\begin{abstract}

In this work, we study the thermodynamic functions of quantum gases confined to spaces of various shapes, namely,
a sphere, a cylinder, and an ellipsoid. We start with the simplest situation, namely, a spinless gas treated within the canonical ensemble framework. As a next step, we consider \textit{noninteracting} gases (fermions and bosons) with the usage of the grand canonical ensemble description. For this case, the calculations are performed numerically. We also observe that our results may possibly be applied to \textit{Bose-Einstein condensate} and to \textit{helium dimer}. Moreover, the bosonic sector, independently of the geometry, acquires entropy and internal energy greater than for the fermionic case. Finally, we also devise a model allowing us to perform analytically the calculations in the case of \textit{interacting} quantum gases, and, afterwards, we apply it to a cubical box.

\end{abstract}

\maketitle

\affiliation{Universidade Federal do Maranh\~{a}o (UFMA), Departamento de
F\'{\i}sica, Campus Universit\'{a}rio do Bacanga, S\~{a}o Lu\'{\i}s - MA,
65080-805, Brazil}
\affiliation{Universidade Estadual do Maranh\~{a}o (UEMA), Departamento de
F\'{i}sica,\\ Cidade Universit\'{a}ria Paulo VI, S\~{a}o Lu\'{i}s - MA,
65055-310, Brazil.}

\section{Introduction}

The investigation of thermal aspects of materials has gained considerable attention in recent years especially in the context of condensed-matter physics and the development of new materials \cite{i1,i8,i2,i3}. Given the existence of some well-known approximations, the electrons of a metal can be assumed to be a gas, as they are effectively free particles \cite{lee1988development,araujo2017structural,aa1,aa2,aa3,aa4}. Such electron systems are worth exploring due to their relevance in fundamental \cite{i5,i7} and applied \cite{i4,i6} physical contexts. In parallel, a longstanding issue in quantum mesoscopic systems is how to perform an exact sum over the states of either \textit{interacting} or \textit{noninteracting} particles. Depending on the situation, the boundary effects cannot be neglected; instead, they should be taken into account in order to acquire a better agreement with the experimental results. Moreover, the properties of some systems are assumed to be shape dependent \cite{i10,Dai2003,Dai2004} and sensitive to their topology \cite{t1,t2,t3,ada2,t5}.

From a theoretical viewpoint, a related problem of statistical mechanics is to perform the sum over all accessible quantum states to obtain the physical quantities \cite{pathria1972statistical,landau2013statistical}. Normally, the spectrum of particle states, which are confined in a volume, will be elucidated by the study of boundary effects. Nevertheless, if the particle wavelength is too short in comparison to the characteristic scale of the system under consideration, boundary effects can be overlooked. In previous years, such an assumption was supported by Rayleigh and Jeans in their radiation theory of electromagnetism \cite{zettili2003}. Furthermore, such an involvement also emerged in a purely mathematical context and was rigorously solved afterwards by Weyl \cite{weyl1968}.

In this work, we study how the thermodynamic functions of quantum gases behave within different
shapes, i.e., spherical, cylindrical, and ellipsoidal ones. Additionally, with possible experimental applications in mind, the thermodynamic functions are calculated for both \textit{noninteracting} and \textit{interacting} particles. Therefore, our results may help in the identification and in further studies involving geometry dependent phenomena within condensed-matter physics and statistical mechanics.

This paper is organized as follows. Initially, in Section \ref{2}, we present a discussion involving the spectral energy for different geometries. Afterwards, in Section \ref{3}, we focus on spinless particles using a setting of the canonical ensemble. Next, in Section \ref{Sec:BF}, we focus on \textit{noninteracting} gases (fermions and bosons) within the same geometries, using the grand canonical ensemble instead. Moreover, in Section \ref{6}, we propose two possible applications of our results, namely, the \textit{Bose-Einstein condensate} and the \textit{helium dimer}. Next, in Section \ref{Sec:Interaction}, we devise a model to perform the calculation of \textit{interacting} quantum gases which is applied to a cubical box. In contrast to the previous sections, the results are derived analytically. Finally, in Section \ref{conclusion}, we conclude and discuss future perspectives.


\section{Spectral energy for different geometries}\label{2}

Initially, we study the thermodynamic properties of confined gases consisting of spinless particles. Moreover, fermions and bosons with a nonzero spin are also taken into account in our investigation. To do so, we must solve the Schr\"{o}dinger equation for particular symmetries with appropriate boundary conditions. With this, the spectral energy can be derived after some algebraic manipulations. In particular, we choose three different geometries, namely, spherical, cylindrical, and ellipsoidal ones. The potentials
for each configuration are given below:
\begin{subequations}
\begin{align}
\mathcal{V}_{\mathrm{Sphere}}\left( r\right) &=\left\{
\begin{array}{c}
0,\text{ if }r<a \\
\infty ,\text{ if }r>a%
\end{array}%
\right. , \label{sphe}
\\
\mathcal{V}_{\text{\textrm{Cylinder}}}\left( \rho,z \right) &=\left\{
\begin{array}{c}
0,\text{ if }\rho <b\text{ and }0<z<2c\text{ } \\
\infty ,\text{otherwise}%
\end{array}%
\right. , \label{cyli}
\\
\mathcal{V}_{\text{\textrm{Ellipsoid}}}\left( x,y,z \right) &=\left\{
\begin{array}{c}
0,\text{ if }x,y,z\text{ satisfy }\frac{x^{2}+y^{2}}{b^{2}}+\frac{z^{2}}{%
c^{2}}<1 \\
\infty ,\text{ if }x,y,z\text{ satisfy }\frac{x^{2}+y^{2}}{b^{2}}+\frac{z^{2}%
}{c^{2}}\geq 1%
\end{array}%
\right. , \label{ellip}
\end{align}
\end{subequations}
where $a$, $b$ and $c$ are geometric parameters defining the size of the potential. In order to make a comparison between our thermodynamic results in the next sections, we must choose the parameters $a$, $b$, and $c$ such that all configurations provide the same volume. Since we have already set up our potentials, we can solve the time-independent Schr\"{o}dinger equation
\begin{equation}
-\frac{\hbar ^{2}}{2\mathrm{M}}\nabla ^{2} \psi+V\left( r\right) \psi =E\psi ,
\end{equation}%
for each geometry whose solutions can be obtained using the well-known method of separation of
variables \cite{Griffthis,ellipsoidal,dirac2001lectures,zettili2003}. Particularly, the wavefunction for the spherical case can be written as 
\begin{equation}
\psi_{l,n,m}\left( r,\theta,\phi\right) =\left\{
\begin{array}{c}
A_{l,n}j_{l}(k_{l,n} r) Y^{m}_{l}(\theta,\phi),\text{ if }r\leq a \\
0 ,\,\,\,\,\,\,\,\,\,\,\,\,\,\,\,\,\,\,\,\,\,\,\,\,\,\,\,\,\,\,\,\,\,\,\,\,\,\,\,\,\,\,\,\,\,\,\,\,\,\text{ if }r \geq a%
\end{array}%
\right. ,
\end{equation}    
where $j_{l}(k_{l,n}r)$ is a spherical Bessel function, $Y^{m}_{l}(\theta,\phi)$ is a spherical harmonic and $A_{l,n}$ is a normalization factor; thereby, the Fourier transform of $\psi_{l,n,m}\left( r,\theta,\phi\right)$ is 
\begin{equation}
\tilde{\psi}_{l,n,m}(k)=   \frac{1}{\sqrt{(2\pi \hbar)^{3}}}  A_{l,n}\int e^{i {\bf{k}}\cdot {\bf{r}}} j_{l}( k_{l,n}r)Y^{m}_{l}(\theta,\phi) r^{2} \mathrm{d}r\mathrm{d}\Omega
\end{equation}
and using the orthogonality properties of the Bessel functions \cite{andrews,silverman,bell}, we can infer the momentum distribution $\overset{\nsim}{n}(k)$
\begin{equation}
\overset{\nsim}{n}(k) = s \sum_{l,n} |\tilde{\psi}_{l,n,m}(k)|^{2} = \frac{3 s V}{4\pi^{2}(\pi\hbar)^{3}} \sum_{l,n} (2l+1) j^{2}_{l}(ka) \frac{(k_{l,n}a/\pi)^{2}}{[(k_{l,n}a/\pi)^{2}- (ka/\pi)^{2}]^{2}},
\end{equation}
where $s$ is the spin degeneracy, $k_{l,n} a$ is the $n$-th root of $j_{l}(ka)=0$, $V$ is the volume, and $k$ the momentum -- it is considered to be continuous in the infinite volume limit. In Ref. \cite{krivine1986}, some approximations in order to perform analytical and numerical analyses of \textit{noninteracting} particles at zero temperature were made. The shape dependence was investigated as well to see how such geometries would influence the momentum distribution $\overset{\nsim}{n}(k)$. Our approach, on the other hand, intends to examine the impact of all mentioned shapes on the thermodynamic properties in comparison to the literature \cite{Dai2003,Dai2004}. To perform the following calculations, we consider spherical, cylindrical, and ellipsoidal configurations. Solving the Schr\"{o}dinger equation for the spherical potential \cite{Griffthis}, we have
\begin{subequations}
\begin{equation}
E_{n,l}^{\text{\textrm{Sphere}}}=\left( 2l+1\right) \frac{\hbar ^{2}}{2\mathrm{M}a^{2}%
}j_{nl}^{2}
\end{equation}%
where $j_{nl}$ is the $n$-th zero of the $l$-th spherical Bessel function.
Clearly, each level has a $\left( 2l+1\right)$-fold degeneracy. For ellipsoidal \cite{ellipsoidal}, cylindrical \cite{Griffthis} shapes, we obtain%
\begin{align}
E_{m,n,l}^{\text{\textrm{Ellipsoid}}}&=\frac{\hbar ^{2}}{2\mathrm{M}}\left[ \frac{J_{n,l}^{2}}{b^{2}}+\frac{2 J_{n,l}}{bc}\left( m+\frac{1}{2}\right) \right]
,
\\
E_{m,n,l}^{\text{\textrm{Cylinder}}}&=\frac{\hbar ^{2}}{2\mathrm{M}}\left[ \frac{J_{n,l}^{2}}{b^{2}}+\left( \frac{\pi m}{2c}\right) ^{2}\right].
\end{align}
\end{subequations}
where $J_{n,l}$ is the $n$-th root of the $l$-th Bessel function, and ${n,m}=0,1,\ldots $. Using these spectral energies, an analysis of the thermal properties can properly be carried out in the next sections.

\section{Noninteracting gases: spinless particles} \label{3}

In this section, we present the thermodynamic approach based on the canonical ensemble.

\subsection{Thermodynamic approach}

Whenever we are dealing with a spinless gas, the theory of the canonical ensemble is sufficient for a full thermodynamical description. Thereby, the partition function is given by%
\begin{equation}
\mathcal{Z}=\sum_{\left\{ \Omega \right\} }\exp \left( -\beta E_{\Omega
}\right) ,  \label{eq:Partition-function}
\end{equation}%
where $\Omega $ is related to accessible quantum states. Since we are dealing with \textit{noninteracting} particles, the partition function $\left( \ref%
{eq:Partition-function}\right)$ can be factorized which gives rise to the result below%
\begin{equation}
\mathcal{Z}=\mathcal{Z}_{1}^{N}=\left\{ \sum_{\left\{ \Omega \right\} }\exp
\left( -\beta E_{\Omega }\right) \right\} ^{N},
\label{eq:Partition-function-1}
\end{equation}%
where we have defined the single partition function as
\begin{equation}
\mathcal{Z}_{1}=\sum_{\Omega }\exp \left( -\beta E_{\Omega }\right) .
\label{eq:Single-Partition-function}
\end{equation}%
It is known that the thermodynamical description of the system can also be done via
Helmholtz free energy%
\begin{equation}
f=-\frac{1}{\beta }\lim_{N\rightarrow \infty }\frac{1}{N}\ln \mathcal{Z},
\end{equation}%
where rather for convenience we write the Helmholtz free energy per particle. With this, we can derive the following thermodynamic state functions\footnote{All of them are written in a “per particle form” for convenience.}, i.e., entropy, heat capacity, and mean energy%
\begin{subequations}

\begin{equation}
s=-\frac{\partial f}{\partial T},  \label{eq:Entropy}
\end{equation}%
\begin{equation}
c=T\frac{\partial s}{\partial T},  \label{eq:heat-capacity}
\end{equation}%
and%
\begin{equation}
u=-\frac{\partial }{\partial \beta }\ln \mathcal{Z}.  \label{eq:Energy}
\end{equation}
\end{subequations}%
The sum in Eq.~(\ref{eq:Single-Partition-function}) cannot be expressed in a closed form. This does not allow us to proceed analytically. To overcome this issue, we perform a numerical analysis plotting the respective functions in terms of the temperature in order to understand their behaviors. Our main interest lies in the study of the low-temperature regime.

\subsection{Numerical analysis}\label{Sec:Numerical1}

\begin{figure}[tbh]
  \centering
  \includegraphics[width=8cm,height=5cm]{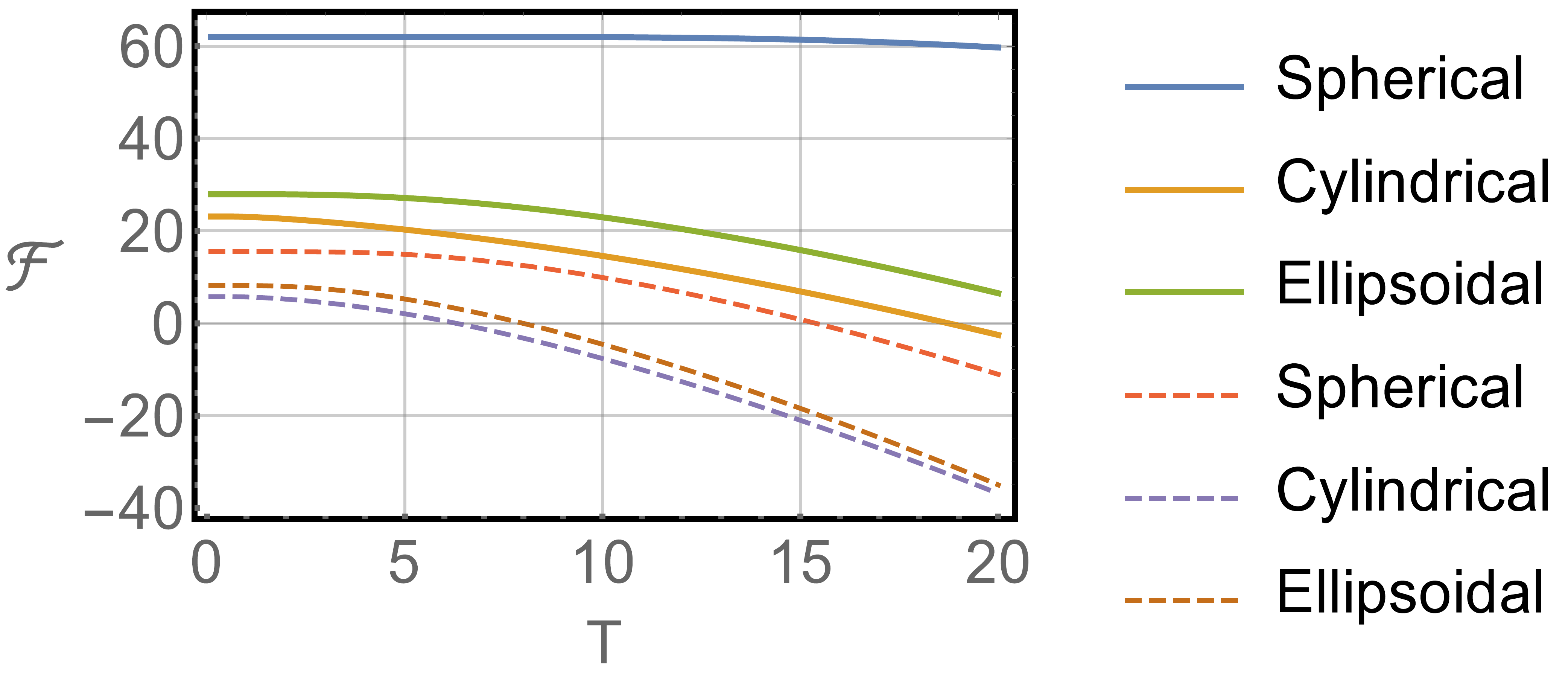}
  \includegraphics[width=8cm,height=5cm]{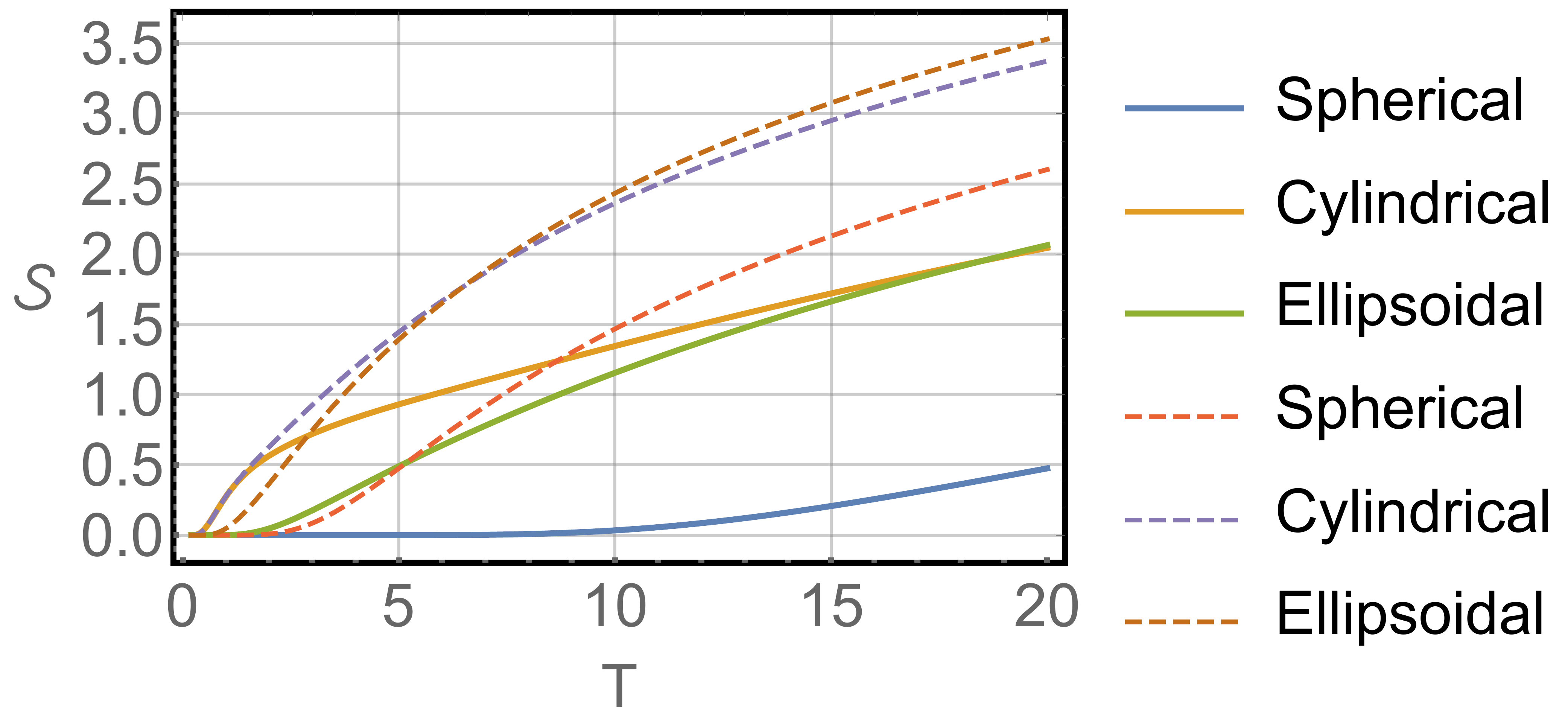}\\
  \includegraphics[width=8cm,height=5cm]{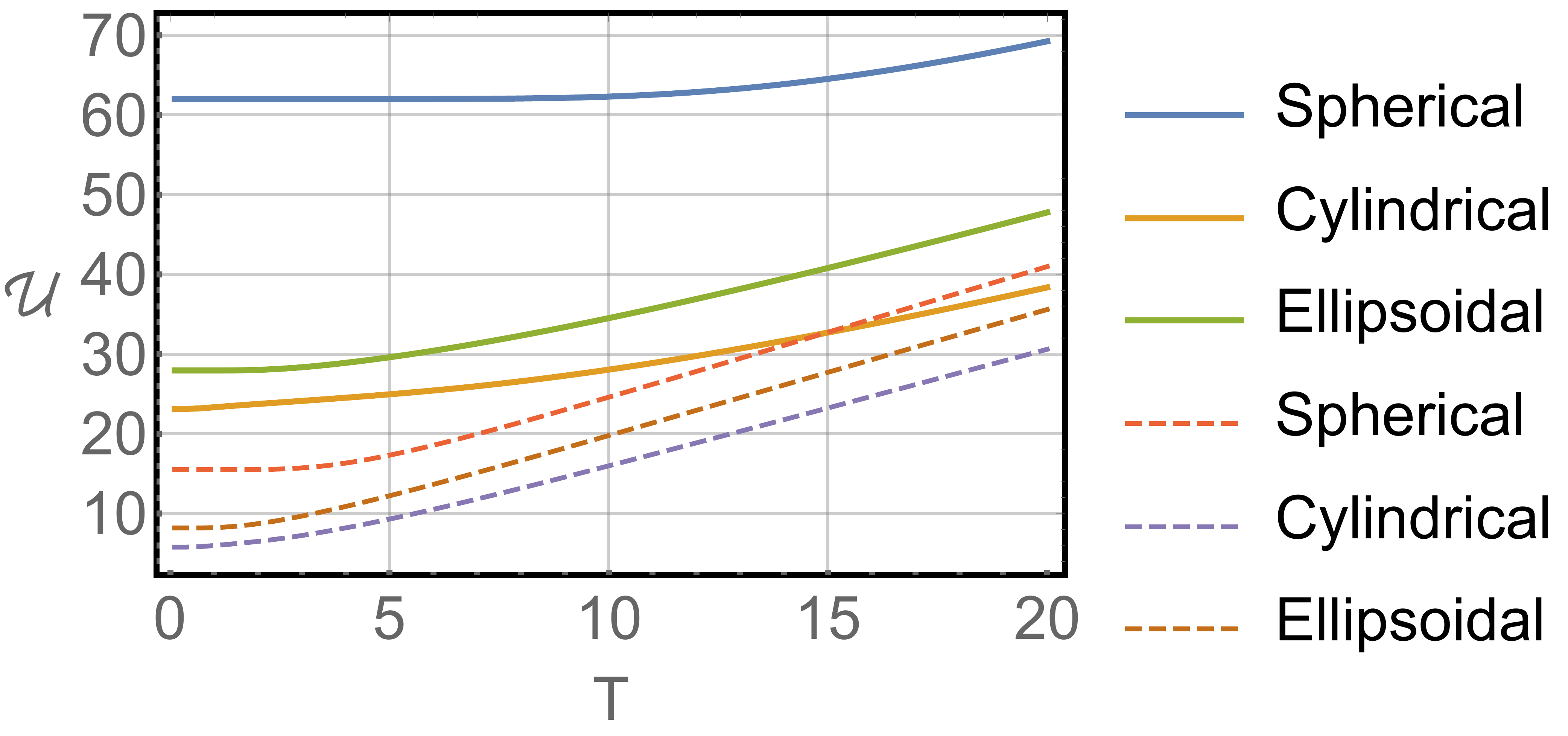}
  \includegraphics[width=8cm,height=5cm]{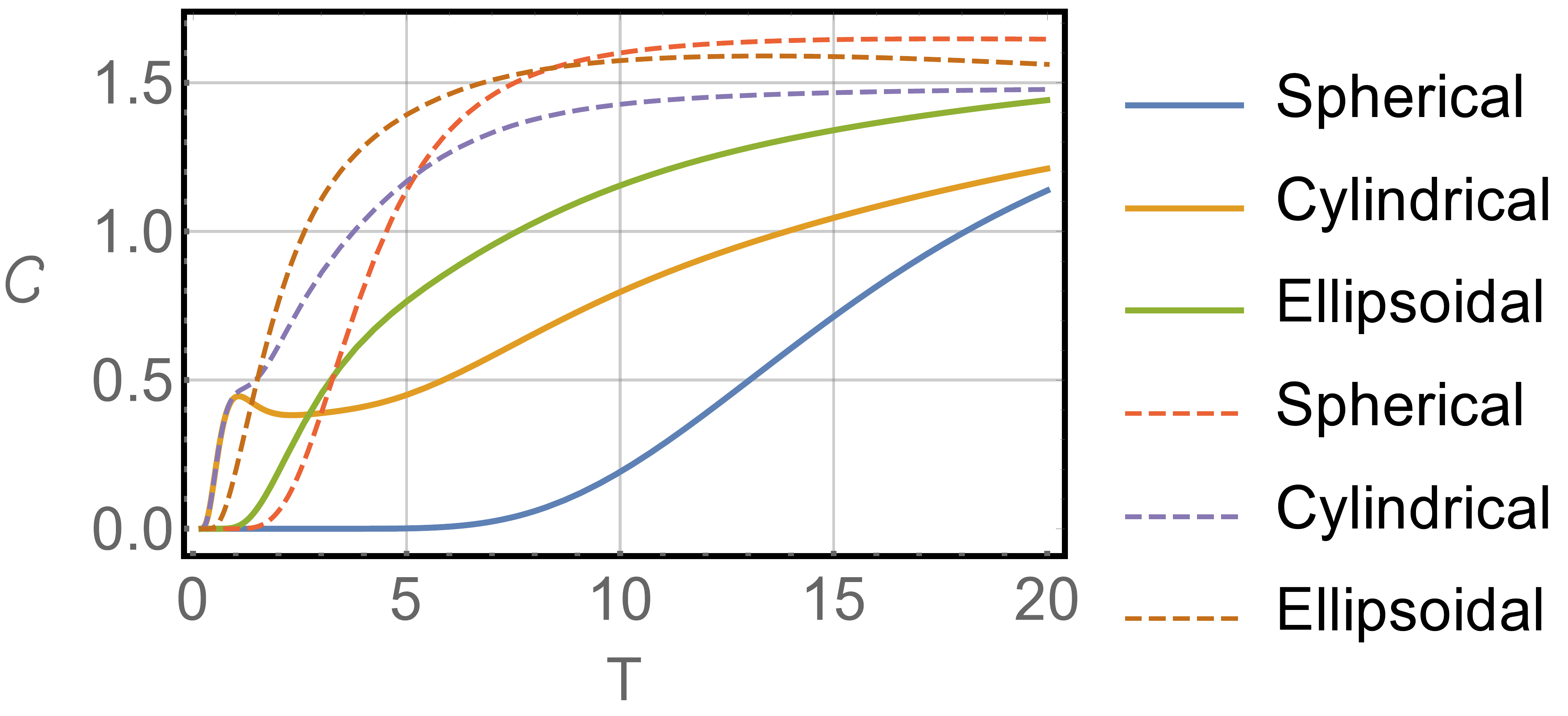}
  \caption{The Helmholtz free energy, entropy per particle, the internal energy and heat capacity per particle, respectively}
  \label{fig:Spin-lessT-C1}
\end{figure}

To provide the numerical results below, we sum over fifty thousand terms in Eq. (\ref{eq:Single-Partition-function}). With this, we can guarantee the accuracy of such procedure keeping the maximum of details. For the plots below, the values for the parameters $a$, $b$ and $c$\footnote{The unit chosen for these parameters in the table is the nanometer. In units of the reduced Compton wavelength, we must remember the conversion factor $1~\mathrm{nm}=2.58\times 10^{3}\lambdabar$. With this information, it is easy to perform the conversion between these units.}, that control the size of the potential, are displayed in Tab. \ref{tab:TableParameters}. For now on, we will refer to the first set of parameters as configuration 1 (thick lines in the plots) and to the second set as configuration 2 (dashed lines in the plots). Note that configuration 2 has a larger volume than configuration 1. However, the parameters are chosen such that the wells considered (sphere, cylinder and ellipsoid) have the same volume in each configuration. We also need to mention that we are using the mass of the electron $M=0.5~\mathrm{MeV}$ in order to provide the numerical results. The graphics for the configurations described here are displayed in Fig. \ref{fig:Spin-lessT-C1}

\begin{table}
  \centering

\begin{tabular}{|c|l|c|c|c|}
  \hline
                              &           & $a$   & $b$  & $c$   \\ \hline\hline
\multirow{3}{5em}{Config. 1}  & Sphere    & 1.0   & -    & -     \\
                              & Cylinder  & -     & 1.0  & 0.66   \\
                              & Ellipsoid & -     & 1.22 & 0.66  \\ [0.5ex] \hline \hline
\multirow{3}{5em}{Config. 2}  & Sphere    & 1.5   & -    & -     \\
                              & Cylinder  & -     & 1.22 & 1.5   \\
                              & Ellipsoid & -     & 1.67 & 1.2   \\
  \hline\hline
\end{tabular}

  \caption{The parameters $a$, $b$ and $c$ for two distinct configurations.}\label{tab:TableParameters}
\end{table}          

Let us start by analyzing the Helmholtz free energy. In Fig. \ref{fig:Spin-lessT-C1}, we see that configuration 1 provides larger values of the Helmholtz free energy than configuration 2. Besides, we also see that the values of the energy are larger for the sphere in comparison to the ellipsoid; and the latter one is larger in comparison to the cylinder for both configurations. In what follows, we will employ the short-hand notation $\textit{Sphere}>\textit{Ellipsoid}>\textit{Cylinder}$. The behavior of the entropy, on the other hand, seems to be different. We notice that configuration 2 provides an entropy larger than configuration 1 and, when we look at the geometry itself, we observe the pattern $\textit{Cylinder}>\textit{Ellipsoid}>\textit{Sphere}$. In this comparison, the \textit{Ellipsoidal} geometry always takes a position in the middle. This fact is expected since this geometry is actually a ``transition'' between a cylinder and a sphere.

Also in Fig. \ref{fig:Spin-lessT-C1}, we find the plots for the internal energy and for the heat capacity. Configuration 1 exhibits energies larger than those of configuration 2 and we see that the internal energy increases according to $\textit{Sphere}>\textit{Ellipsoid}>\textit{Cylinder}$ for both configurations. The heat capacities for all configurations approach the value $3/2$ as the temperature increases. Configuration 2, that has a larger volume, reaches the asymptotic value faster than configuration 1.

Differently from what we saw for the other thermodynamic quantities, it is not possible to establish a common behavior among \textit{Ellipsoid}, \textit{Cylinder} and \textit{Sphere} geometries because of different temperature ranges and well sizes the heat capacity varies drastically. For instance, in the second row of Fig. \ref{fig:Spin-lessT-C1}, configuration 2 in the range $0<T<5$~K, we see that the heat capacity follows the rule $\textit{Ellipsoid}>\textit{Cylinder}>\textit{Sphere}$. However, in the same range of temperature, configuration 1 displays two different behaviors, i.e., for $0<T<2.5$~K,  we have $\textit{Cylinder}>\textit{Ellipsoid}>\textit{Sphere}$ and for $2.5$~K $<T<5$~K, we have $\textit{Ellipsoid}>\textit{Cylinder}>\textit{Sphere}$. For a fixed value of the volume, there exists a temperature where the heat capacity will follow the rule $\textit{Sphere}>\textit{Ellipsoid}>\textit{Cylinder}$ until it reaches the value $3/2$. This temperature increases as the volume decreases. For configuration 1, the temperature where we have this behavior is around $29$~K and, for configuration 2, it happens at $9$~K.

It is interesting to see that for the cylindrical geometry a mound appears around $0.5$~K and tends to disappear when the volume of the cylinder increases. This effect is clearly caused by the finite size of the geometry since the expected behavior would be to go to zero almost linearly. However, both the sphere and the ellipsoid do not exhibit such an effect. We could conclude about the absence of such effect that the surfaces of the former cases that accentuate those geometries are smooth and the cylinder, on the other hand, is smooth just by parts. As we will see in Sec. \ref{Sec:Interaction} and in Ref. \cite{Dai2004}, it is possible to identify the contribution that comes from the geometry itself by considering an analytical model.

\section{Noninteracting gases: Bosons and fermions }\label{Sec:BF}

Although interactions of atoms and molecules are treated in many experimental approaches, and several features may only be recognized and understood by taking the interactions into account \cite{i11,i22,i33,i44}, some fascinating characteristics are well described by assuming \textit{noninteracting} systems  \cite{chen1972light,e1,e2,e3,e4,e5,e6,e7,e8,e9,e10,e11,e12}.

Studies of \textit{noninteracting} particles (bosons and fermions) have many applications, especially in chemistry \cite{e10,e11} and condensed-matter physics \cite{chen1972light,e1,e2,e3,e4,e5,e6,e7,e8,e9,e12,boseeinteincondensate2}; for instance, in the case of \textit{bulk}, which is usually assumed to calculate the energy spectrum and use the \textit{Fermi-Dirac} distribution to examine how its statistics behaves, it is sufficient to describe the system of a \textit{noninteracting} electron gas. Such an assumption is totally reasonable since if the \textit{Fermi energy} is large enough, the kinetic energy of electrons, close to the Fermi level, will be much greater than the potential energy of the \textit{electron-electron} interaction.

On the other hand, one of the pioneer studies which addressed the analysis of the \textit{Bose-Einstein condensate} in a theoretical viewpoint was presented in \cite{pajkowski1977}. The authors utilize a gas of \textit{noninteracting} bosons to perform their calculations. As we shall see, we proceed in a similar way taking into account different geometries. Furthermore, we discuss applying them in different scenarios in condensed-matter physics.

\subsection{Thermodynamic approach}

We apply the grand canonical ensemble theory to \textit{noninteracting} particles with different spins (fermions and bosons); we will treat both cases separately. The grand canonical partition function for the present problem reads%
\begin{equation}
\Xi =\sum_{N=0}^{\infty }\exp \left( \beta \mu N\right) \mathcal{Z}\left[
N_{\Omega }\right] ,  \label{eq:GarndPartition-function}
\end{equation}%
where $\mathcal{Z}\left[ N_{\Omega }\right] $ is the usual canonical partition function which now depends on the occupation number $N_{\Omega }$, and on $\mu$, which is the chemical potential. Since we are dealing with fermions and bosons, it is well-known that the occupation number must be restricted in the following manner: $N_{\Omega }=\left\{ 0,1\right\} $ for fermions and $N_{\Omega
}=\left\{ 0,\ldots \infty \right\} $ for bosons. Also, for an arbitrary quantum state, the energy depends on the occupation number as
\begin{equation*}
E\left\{ N_{\Omega }\right\} =\sum_{\left\{ \Omega \right\} }N_{\Omega
}E_{\Omega }
\end{equation*}%
where we have%
\begin{equation*}
\sum_{\left\{ \Omega \right\} }N_{\Omega }=N.
\end{equation*}%
In this way, the partition function becomes%
\begin{equation}
\mathcal{Z}\left[ N_{\Omega }\right] =\sum_{\left\{ N_{\Omega }\right\}
}\exp \left[ -\beta \sum_{\left\{ \Omega \right\} }N_{\Omega }E_{\Omega }%
\right],
\end{equation}%
which leads to%
\begin{equation}
\Xi =\sum_{N=0}^{\infty }\exp \left( \beta \mu N\right) \sum_{\left\{
N_{\Omega }\right\} }\exp \left[ -\beta \sum_{\left\{ \Omega \right\}
}N_{\Omega }E_{\Omega }\right] ,
\end{equation}%
or can be rewritten as
\begin{equation}
\Xi =\prod_{\left\{ \Omega \right\} }\left\{ \sum_{\left\{ N_{\Omega
}\right\} }\exp \left[ -\beta N_{\Omega }\left( E_{\Omega }-\mu \right) %
\right] \right\} .
\end{equation}%
After performing the sum over the possible occupation numbers, we get%
\begin{equation}
\Xi =\prod_{\left\{ \Omega \right\} }\left\{ 1+\chi \exp \left[ -\beta
\left( E_{\Omega }-\mu \right) \right] \right\} ^{\chi },
\end{equation}%
where we have now introduced the convenient notation $\chi =+1$ for fermions and $\chi =-1$ for bosons. The connection with thermodynamics is made by using the grand thermodynamical potential given by%
\begin{equation}
\Phi =-\frac{1}{\beta }\ln \Xi .
\end{equation}%
Replacing $\Xi $ in the above equation, we get%
\begin{equation}
\Phi =-\frac{\chi }{\beta }\sum_{\left\{ \Omega \right\} }\ln \left\{ 1+\chi
\exp \left[ -\beta \left( E_{\Omega }-\mu \right) \right] \right\} .
\label{eq:Gand-potential}
\end{equation}%
The entropy of the system can be cast into the following compact form, namely%
\begin{equation*}
S=-\frac{\partial \Phi }{\partial T}=-k_{B}\sum_{\left\{ \Omega \right\} }%
\mathcal{N}_{\Omega }\ln \mathcal{N}_{\Omega }+\chi \left( 1-\chi \mathcal{N}%
_{\Omega }\right) \ln \left( 1-\chi \mathcal{N}_{\Omega }\right)
\end{equation*}%
where%
\begin{equation*}
\mathcal{N}_{\Omega }=\frac{1}{\exp \left[ \beta \left( E_{\Omega }-\mu
\right) \right] +\chi }.
\end{equation*}%
Moreover, we can also use the grand potential to calculate other thermodynamic properties, such as, the mean particle number, energy, heat capacity, and
pressure using the following expressions:%
\begin{subequations}
\begin{eqnarray}
\mathcal{N} &=&-\frac{\partial \Phi }{\partial \mu }, \\
\mathcal{U} &=&-T^{2}\frac{\partial }{\partial T}\left( \frac{\Phi }{T}%
\right) , \\
C &=&T\frac{\partial S}{\partial T}, \\
\mathcal{P} &=&-\frac{\partial \Phi }{\partial V}=-\frac{\Phi }{V}.
\end{eqnarray}
\end{subequations}

In possession of these terms, calculating the thermodynamic quantities should be a straightforward task, since we would only need to perform the sum presented in Eq. $\left( \ref{eq:Gand-potential}%
\right) $. Unfortunately, this sum cannot be obtained in a closed form for the spectral energy that we chose.
Instead of this, a numerical analysis, similar to that shown in Sec. \ref{Sec:Numerical1}, can be performed to overcome this difficulty; thereby, we can obtain the behavior of all quantities considering mainly the low temperature regimes (keeping the volume constant). In what follows, we present the results of our numerical studies.


\subsection{Numerical analysis}\label{Sec:Numerical2}

\begin{figure}[tbh]
\centering
\subfloat[Entropy case 1]{
  \includegraphics[width=8cm,height=5cm]{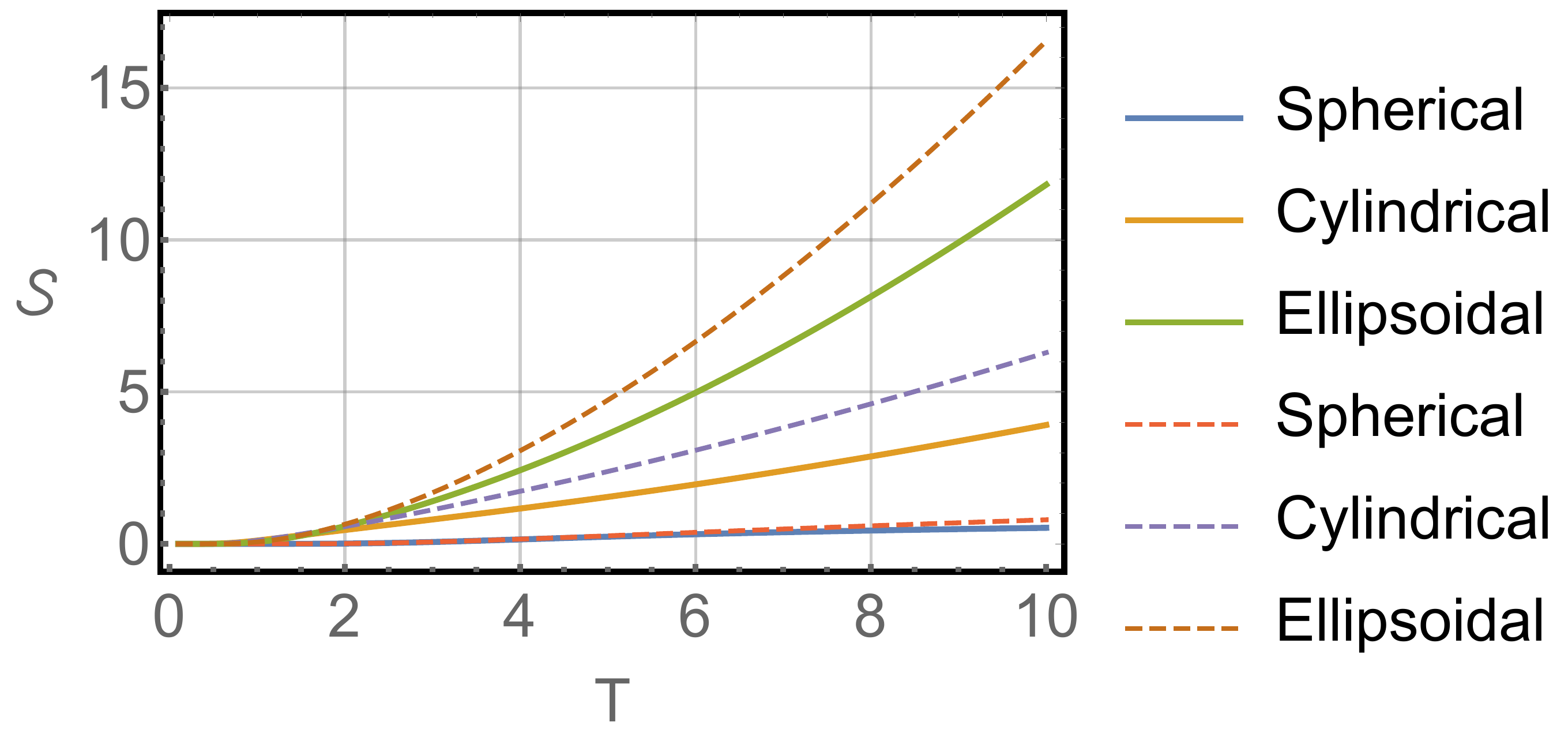}
  \label{fig:S-1-BF}}
\subfloat[Entropy case 2]{
  \includegraphics[width=8cm,height=5cm]{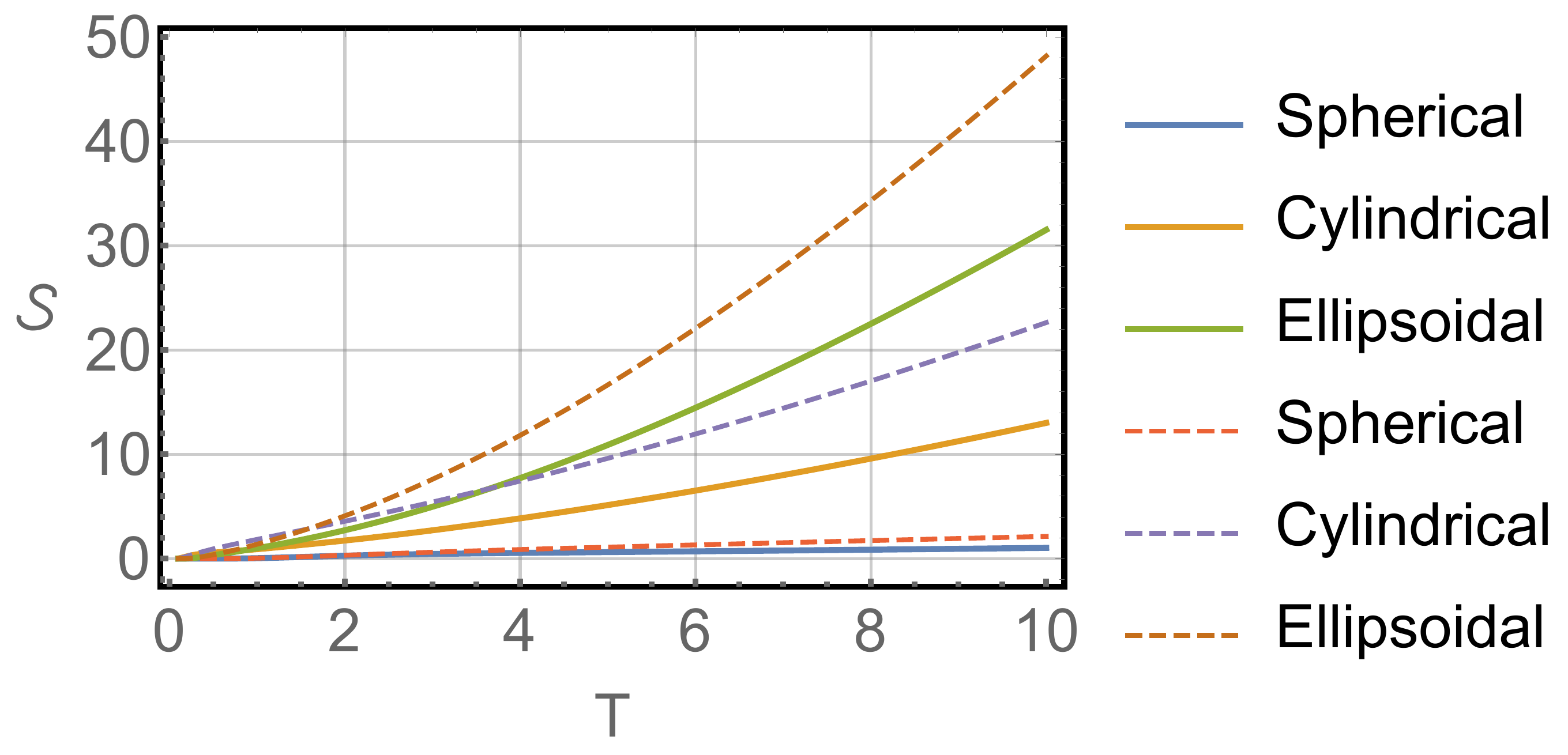}
  \label{fig:S-2-BF}}\\
\subfloat[Energy case 1]{
  \includegraphics[width=8cm,height=5cm]{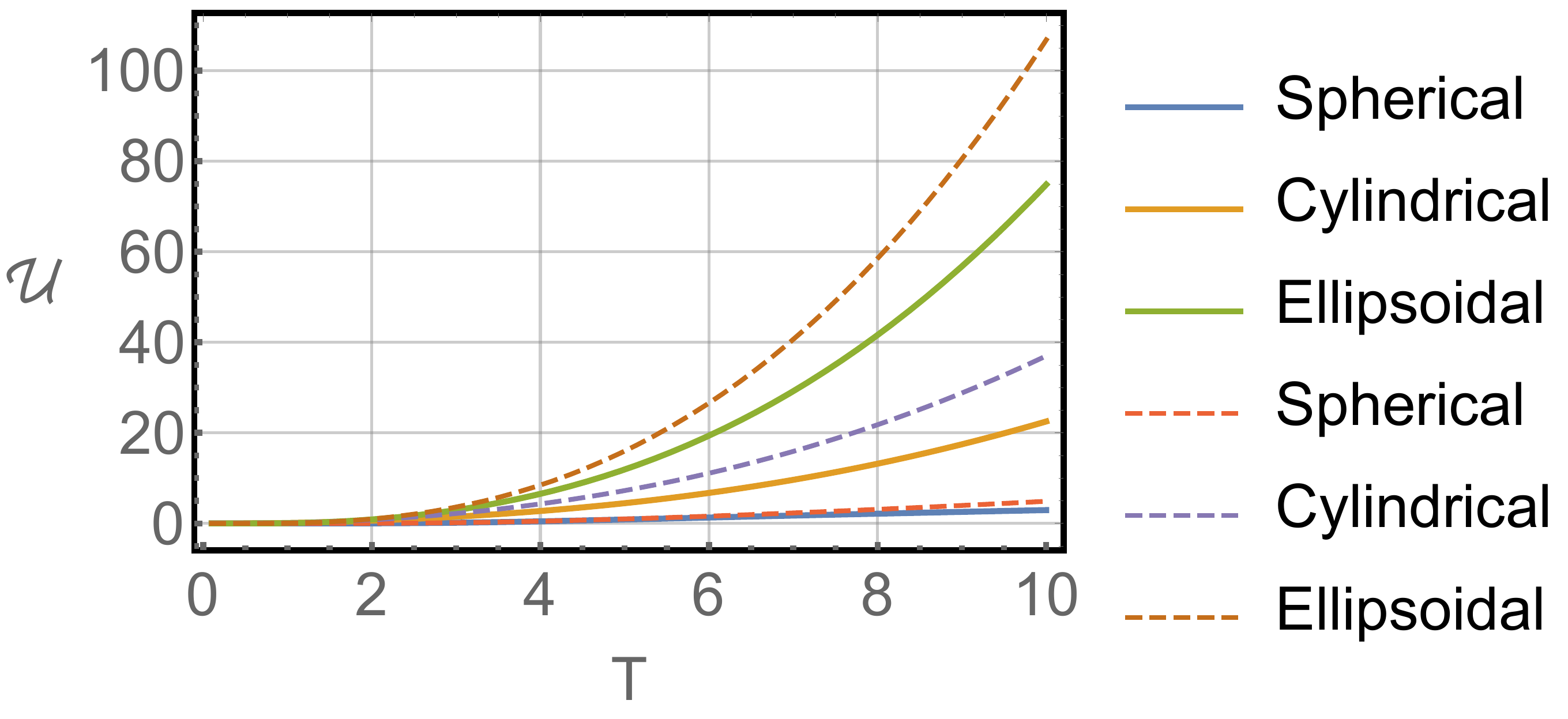}
  \label{fig:E-1-BF}}
\subfloat[Energy case 2]{
  \includegraphics[width=8cm,height=5cm]{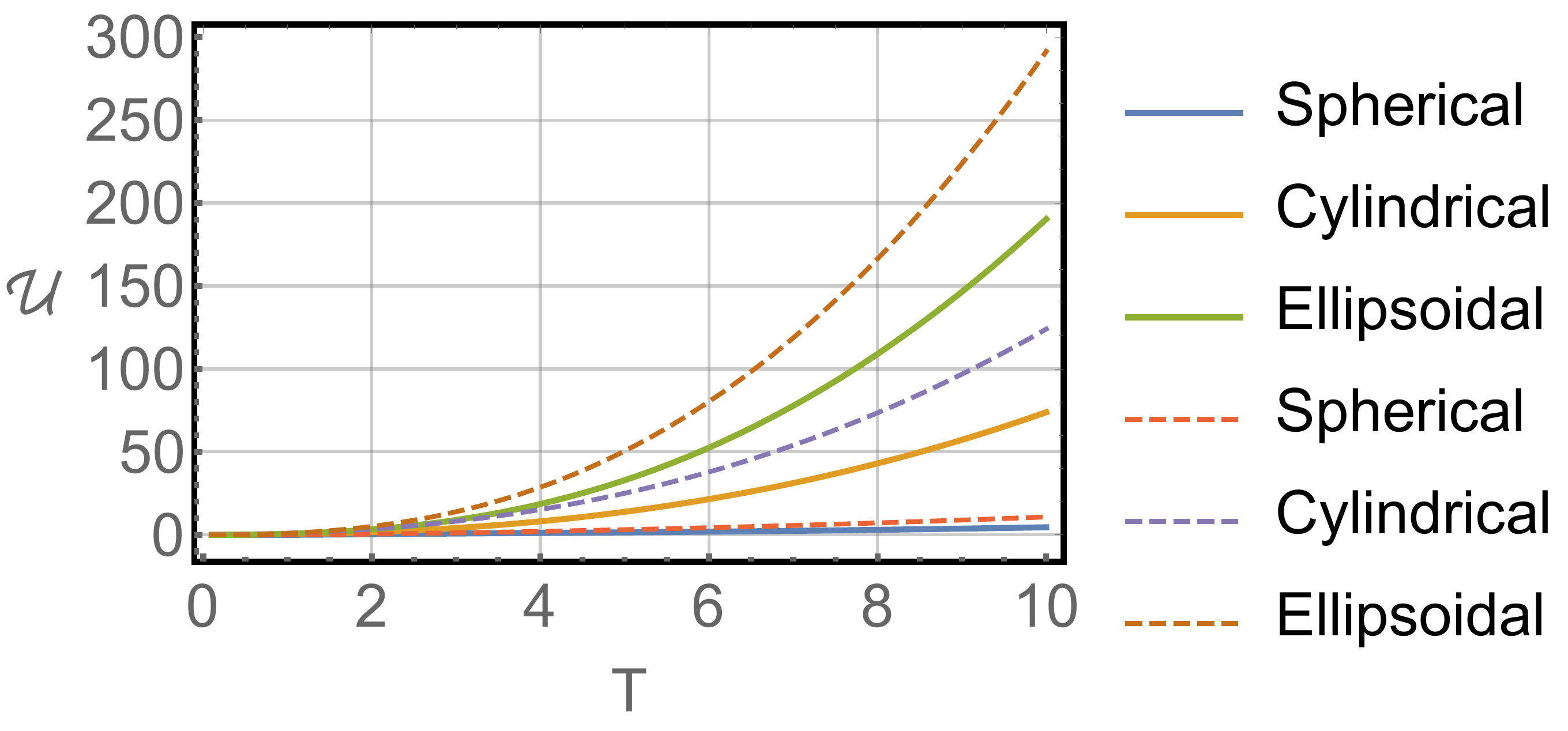}
  \label{fig:E-2-BF}}\\
\subfloat[Heat capacity case 1]{
  \includegraphics[width=8cm,height=5cm]{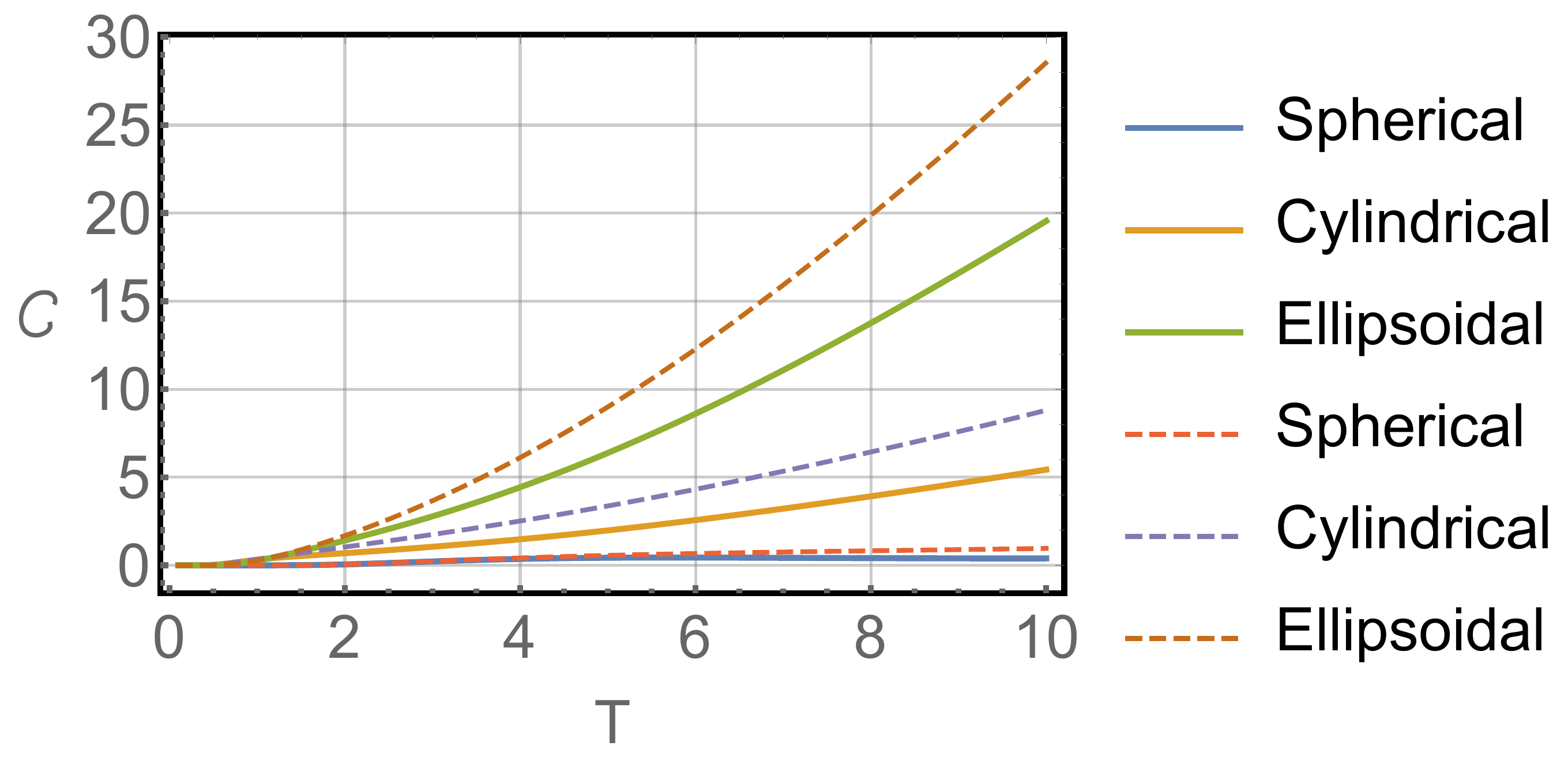}
  \label{fig:C-1-BF}}
\subfloat[Heat capacity case 2]{
  \includegraphics[width=8cm,height=5cm]{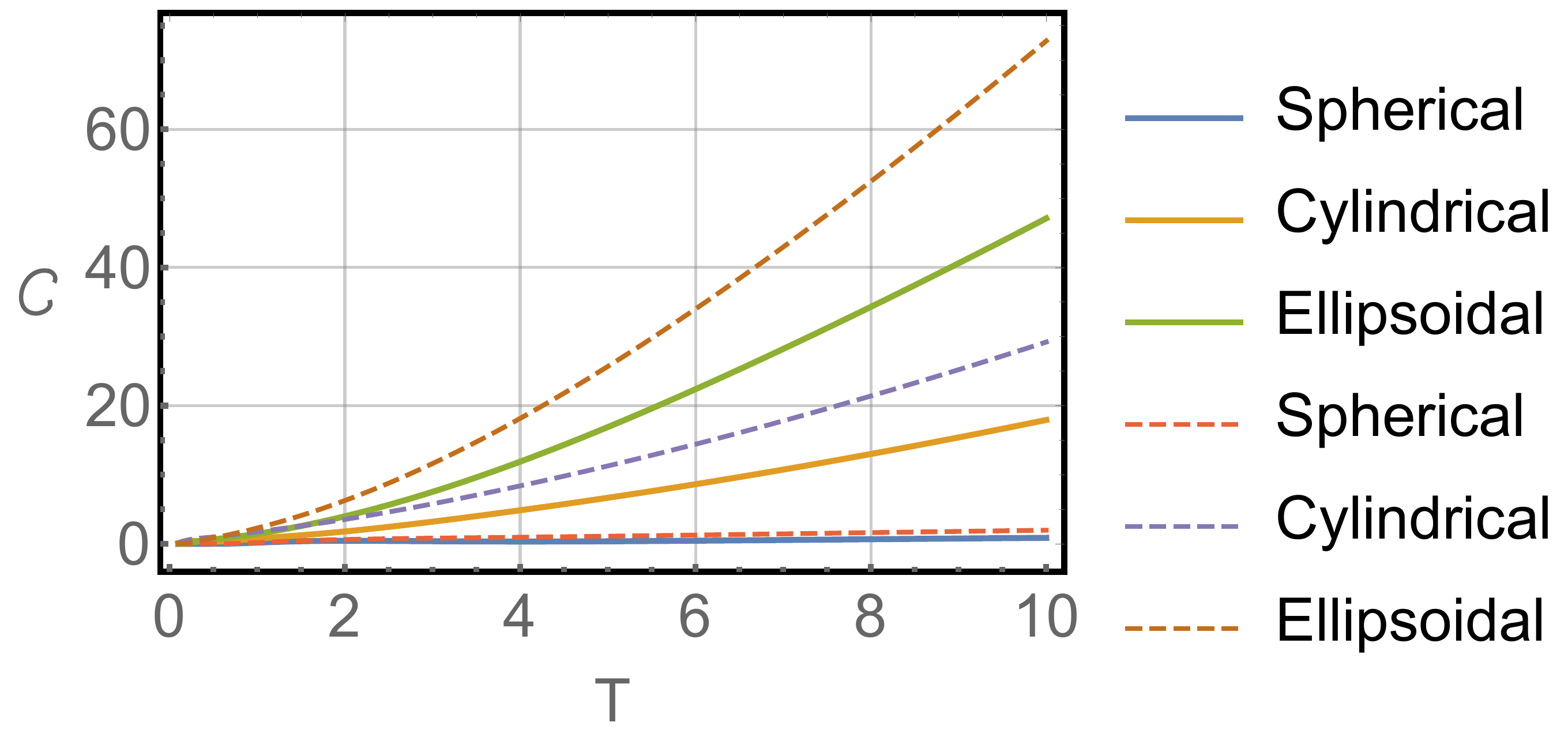}
  \label{fig:C-2-BF}}  
\caption{The different behaviors for the entropy, mean energy and heat capacity, respectively}
\label{fig:BF-S}
\end{figure}

Here, as in Section \ref{Sec:Numerical1}, we consider two particular cases whose parameters are the same as those presented in Tab. \ref{tab:TableParameters}. For numerical purpose, we choose for the chemical potential the value $\mu = 0.5~\mathrm{eV}$ and, again, $M=0.5~\mathrm{MeV}$. Next, we also use thick and dashed lines to represent fermions and bosons in the plots presented in Fig. \ref{fig:BF-S}. In the first row of Fig. \ref{fig:BF-S}, we present the entropy for the two different cases. The first one, in Fig. \ref{fig:S-1-BF}, represents configuration 1 and Fig. \ref{fig:S-2-BF} configuration 2, where we compare fermions (thick lines) and bosons (dashed lines) for sphere, cylinder and ellipsoid geometries. We can see from the plots that bosons, independently of the previous set of geometry we chose, acquire an entropy larger than fermions. We can also realize that the pattern $\textit{Ellipsoid}>\textit{Cylinder}>\textit{Sphere}$ always occurs for fermions and bosons when we consider the entropy. In both cases, the pattern described above repeats and shows that the entropy is a monotonically increasing function for volume and temperature.

For the internal energy, displayed in the second row Fig. \ref{fig:BF-S}, we see the entire behavior being repeated, namely, the internal energy follows the rule $\textit{Ellipsoid}>\textit{Cylinder}>\textit{Sphere}$ and the energy for bosons are greater than for fermions when we compare their values for the same geometry. It is also important to notice that the internal energy is a monotonically increasing function for both volume and temperature.

Another important property analyzed here is the heat capacity showed in the third row of Fig. \ref{fig:BF-S}. We know from the literature \cite{pathria1972statistical,landau2013statistical} that the heat capacity for an electron gas at low temperature ($T\ll T_{Fermi}$) is proportional to the temperature. On the other hand, for the boson gas, the heat capacity is proportional to $T^{3/2}$ (both behaviors are obtained in the classical limit). However, as we can deduce from the graphics, in both cases presented in Fig. \ref{fig:BF-S}, fermions do not follow this rule. We see deviations form a straight line, where one can infer that this effect is caused by the finite volume. We must remember that our systems are under confinement in very small containers\footnote{The linear dimension of the containers considered is comparable with the thermal wavelength of the electron for low temperatures. This means that the electrons will feel the finiteness of the well and, therefore, the thermodynamic properties will change.} and, in addition, a large number of energy levels  were considered in the numerical analysis. As a direct consequence, the entire behavior at low temperature, when compared to the usual electron gas, will be drastically different. On the other hand, bosons behave exactly like a function proportional to $T^{3/2}$. It suggests that fermions are more sensitive to the geometry and size of the system than bosons. Besides the discussion above, we can also see that in general the heat capacity follows the pattern $\textit{Ellipsoid}>\textit{Cylinder}>\textit{Sphere}$ and its value increases when both temperature and volume increase.

\section{Further applications: noninteracting gases}\label{6}

In this section, we address qualitatively some possible future applications of our \textit{noninteracting} model for quantum gases developed so far. In particular, we look toward the \textit{Bose-Einstein condensate} and the \textit{helium dimer}.

\subsection{Bose-Einstein condensate}

In Ref. \cite{pajkowski1977}, the grand canonical ensemble is also used to perform the calculations; the authors studied the asymptotic behavior of various thermodynamic and statistical quantities related to a confined ideal \textit{Bose-Einstein} gas. In this case, the considered object is an arbitrary, finite, cubical enclosure subjected to periodic boundary conditions, i.e., thin-film, square-channel and cubic geometries.

Following the approach of Ref. \cite{pajkowski1977}, our proposition is to probe how the thermal quantities are affected by spherical, cylindrical, and ellipsoidal geometries. This study can be useful within a possible future experimental scenario to be studied in material science.

\subsection{Helium atoms - $^{3}\mathrm{He}$ and $^{4}\mathrm{He}$ }

Taking the advantage of solving the Schrödinger equation numerically as well as the construction of suitable wavefunctions, in Refs. \cite{kilic1999,kilic2000}, the binding of two helium atoms involving restricted and unrestricted geometries was studied in two and three dimensions.
Such a model describes two atoms placed in a spherical potential (\ref{sphe}) with hard walls. As argued, one could insert a nontrivial interaction of the helium atoms with the walls \cite{gersbacher1972,zaremba,bretz1971,bretz19711} and also some coupling between them, as presented in Sec. \ref{Sec:Interaction}. Nevertheless, the interaction of the particles with the wall depends on the material of the cavity. Thereby, it is feasible to propose a general investigation of these phenomena rather than being limited to individual cases. 

In this sense, our proposal is as follows: 
based on the relevance of studying either helium liquids or helium dimer in solid matrices, a study of how geometry influences the thermodynamic properties of such constrained systems might be relevant for future applications in condensed matter physics. Likewise, for the cylindrical shape (\ref{cyli}), it is notable to aim at investigating the thermal properties of shapes similar to that of carbon nanotubes \cite{dresselhaus2000carbon,jorio2007carbon}. Also, using some approximations, vortex-like shapes \cite{saarela1993phase,saarela1995many} could reasonably be examined as well.


\section{Interacting gases: an analytical approach}\label{Sec:Interaction}

\subsection{The model}

We intend now to take into account interactions between particles. To do so, we modify slightly the approach developed in Sec. \ref{Sec:BF}
by introducing an interaction term $U\left( V,n\right) $. Here, we assume that the interaction energy depends only on the particle density $n$ and the volume $V$. As demonstrated below, such a type of interaction appears in the mean field approximation, and enables us to derive analytical results. Furthermore, as we shall verify, these solutions will allow us to recognize how the interactions can modify the thermal quantities of our system. It is important to highlight that the interaction term is a monotonically increasing function of the particle density. Thereby, if the density $n$ is increased, the particles will come closer to each other and the respective interactions between them are supposed to increase. Analogously, the opposite behavior happens otherwise: if $n$ decreases, $U\left( V,n\right)$ will have to decrease. Throughout this section, we adopt natural units where $k_{B}=1$. Doing that, we get the following grand canonical partition function 
\begin{subequations}
\begin{equation}
\mathcal{Z}\left( T,V,\mu \right) =\sum_{\left\{ N_{\Omega }\right\}
=0}^{\left\{ \infty /1\right\} } z^{N}\exp \left\{ -\beta \left[ \sum_{\left\{
\Omega \right\} }N_{\Omega }\left( E_{\Omega } \right) +U\left(
V,n\right) \right] \right\} ,  \label{eq:g-partition-function}
\end{equation} 
where
\begin{equation}
z^{N}=\exp \left\{ N\beta \mu \right\} =\exp \left\{ \beta \sum_{\left\{
\Omega \right\} }N_{\Omega }\mu \right\} .
\end{equation}
\end{subequations}%
The sum index which appears
in Eq. (\ref{eq:g-partition-function}), namely $\left\{ \infty /1\right\} $, shows that infinitely many bosons may occupy the same quantum state $\Omega$. On the other hand, if one considers instead of this, spin-half particles, only one fermion will be allowed due to the Pauli exclusion principle. In a compact notation, we take the upper index $\infty $ for bosons and $1$ for fermions. Let us now suppose that the interaction term has the form $U\left( V,n\right) =Vu\left( n\right) $. With this, we have%
\begin{equation}
\mathcal{Z}\left( T,V,\mu \right) =\sum_{\left\{ N_{\Omega }\right\}
=0}^{\left\{ \infty /1\right\} }\exp \left\{ -\beta \left[ \sum_{\left\{
\Omega \right\} }N_{\Omega }\left( E_{\Omega } - \mu \right) +Vu\left(
n\right) \right] \right\} .  \label{eq:g-partition-function-1}
\end{equation}

For the sake of simplicity, in Eq. (\ref{eq:g-partition-function-1}), we assume that $Vu\left( n\right) $ is linear in $\sum_{\Omega
}N_{\Omega }=N$. Furthermore, the only appropriate manner to do so is to linearize $%
Vu\left( n\right) $. To do that, we use the Taylor
series expansion of $u\left( n\right) $ around the mean value $\bar{n}$:%
\begin{equation}
u\left( n\right) =u\left( \bar{n}\right) +u^{\prime }\left( \bar{n}\right)
\left( n-\bar{n}\right) +\ldots. \label{eq:Taylor_u}
\end{equation}%
More so, if one regards that the potential energy is dependent on the position that the particles occupy, Eq. (\ref{eq:Taylor_u}) will account for the \textit{molecular field approximation}. Such assumption is vastly employed in the literature concerning for instance condensed matter physics \cite{humphries1972,klein1969,wojtowicz,ter1962molecular,araujo2017,silva2018}. Thereby, we can derive the energy of the quantum state $\Omega$ as being%
\begin{equation}
E=\sum_{\left\{ \Omega \right\} }N_{\Omega }\left[E_{\Omega }+u^{\prime }\left( \bar{n}\right)\right]+U\left( V,\bar{n}%
\right) -u^{\prime }\left( \bar{n}\right) \bar{N}, \label{eq:Total_energy}
\end{equation}%
and the summation of Eq. (\ref{eq:g-partition-function-1}) can be evaluated%
\begin{eqnarray}
\mathcal{Z}\left( T,V,\mu \right) &=&\exp \left\{ -\beta \left[ U\left( V,%
\bar{n}\right) -u^{\prime }\left( \bar{n}\right) \bar{N}\right] \right\}
\notag \\
&&\times \prod_{\Omega =1}^{\infty }\left( \sum_{N_{\Omega }=0}^{\left\{
\infty /1\right\} }\exp \left\{ -\beta \left[ E_{\Omega }+u^{\prime }\left(
\bar{n}\right) -\mu \right] N_{\Omega }\right\} \right) .
\label{eq:Modified_GCP}
\end{eqnarray}%
After some algebraic manipulations, we can present the above expression as%
\begin{eqnarray}
\mathcal{Z}\left( T,V,\mu \right) &=&\exp \left\{ -\beta \left[ U\left( V,%
\bar{n}\right) -u^{\prime }\left( \bar{n}\right) \bar{N}\right] \right\}
\notag \\
&&\times \prod_{\Omega =1}^{\infty }\left\{
\begin{array}{c}
1+\exp \left[ -\beta \left( E_{\Omega }+u^{\prime }\left( \bar{n}\right)
-\mu \right) \right] ,\text{fermions} \\
\left( 1-\exp \left[ -\beta \left( E_{\Omega }+u^{\prime }\left( \bar{n}%
\right) -\mu \right) \right] \right) ^{-1},\text{bosons}%
\end{array}%
\right. ,
\end{eqnarray}%
or in a more compact form
\begin{eqnarray}
\mathcal{Z}\left( T,V,\mu \right) &=&\exp \left\{ -\beta \left[ U\left( V,%
\bar{n}\right) -u^{\prime }\left( \bar{n}\right) \bar{N}\right] \right\}
\notag \\
&&\times \prod_{\Omega =1}^{\infty }\left( 1+\chi \exp \left[ -\beta \left(
E_{\Omega }+u^{\prime }\left( \bar{n}\right) -\mu \right) \right] \right)
^{\chi },
\end{eqnarray}%
where $\chi=1$ for fermions, and $\chi=-1$ for bosons.

\subsection{Thermodynamic state quantities}

Next, the derivation of the grand canonical potential is straightforward as follows%
\begin{eqnarray}
\Phi &=&-T\ln \mathcal{Z}  \notag \\
&=&-T\chi \sum_{\Omega }\ln \left( 1+\chi \exp \left[ -\beta \left(
E_{\Omega }+u^{\prime }\left( \bar{n}\right) -\mu \right) \right] \right)
+U\left( V,\bar{n}\right) -u^{\prime }\left( \bar{n}\right) \bar{N}.
\label{eq:grand-canonical-potencial}
\end{eqnarray}
Based on this equation, the other thermodynamic functions can be calculated as well. In this sense, the mean particle number reads%
\begin{align}
\bar{N}& =-\left. \frac{\partial \Phi }{\partial \mu }\right\vert _{T,V},%
\displaybreak[0]  \notag \\
& =-V\left. \frac{\partial u\left( \bar{n}\right) }{\partial \mu }%
\right\vert _{T,V}+\bar{N}\left. \frac{\partial u^{\prime }\left( \bar{n}%
\right) }{\partial \mu }\right\vert _{T,V}+u^{\prime }\left( \bar{n}\right)
\left. \frac{\partial \bar{N}}{\partial \mu }\right\vert _{T,V}+%
\displaybreak[0]  \notag \\
& +T\chi \sum_{\Omega }\frac{\chi \exp \left[ -\beta \left( E_{\Omega
}+u^{\prime }\left( \bar{n}\right) -\mu \right) \right] }{1+\chi \exp \left[
-\beta \left( E_{\Omega }+u^{\prime }\left( \bar{n}\right) -\mu \right) %
\right] }\beta \left( 1-\left. \frac{\partial u^{\prime }\left( \bar{n}%
\right) }{\partial \mu }\right\vert _{T,V}\right) ,
\end{align}%
and from this,%
\begin{eqnarray}
\bar{N}\left( 1-\left. \frac{\partial u\left( \bar{n}\right) }{\partial \mu }%
\right\vert _{T,V}\right) &=&-V\left. \frac{\partial u\left( \bar{n}\right)
}{\partial \mu }\right\vert _{T,V}+u^{\prime }\left( \bar{n}\right) \left.
\frac{\partial \bar{N}}{\partial \mu }\right\vert _{T,V}  \\
&&+\chi ^{2}\beta T\left( 1-\left. \frac{\partial u^{\prime }\left( \bar{n}%
\right) }{\partial \mu }\right\vert _{T,V}\right) \sum_{\Omega }\frac{1}{%
\exp \left[ \beta \left( E_{\Omega }+u^{\prime }\left( \bar{n}\right) -\mu
\right) \right] +\chi }. \nonumber \label{eq:Mean-N-1}
\end{eqnarray}%
Since %
\begin{equation}
\left. \frac{\partial u\left( \bar{n}\right) }{\partial \mu }\right\vert
_{T,V}=u^{\prime }\left( \bar{n}\right) \left. \frac{\partial \bar{n}}{%
\partial \mu }\right\vert _{T,V}=\frac{u^{\prime }\left( \bar{n}\right) }{V}%
\left. \frac{\partial \bar{N}}{\partial \mu }\right\vert _{T,V},
\label{eq:relation-u}
\end{equation}%
we get%
\begin{equation}
\bar{N}=\sum_{\Omega }\frac{1}{\exp \left[ \beta \left( E_{\Omega
}+u^{\prime }\left( \bar{n}\right) -\mu \right) \right] +\chi }.
\label{eq:Mean-number-N}
\end{equation}%
The mean occupation number
must be $\bar{N}=\sum_{\Omega }\bar{n}_{\Omega }$, where%
\begin{equation}
\bar{n}_{\Omega }=\frac{1}{\exp \left[ \beta \left( E_{\Omega }+u^{\prime
}\left( \bar{n}\right) -\mu \right) \right] +\chi }.
\label{eq:Mean-number-N-new-n1}
\end{equation}
As we can also notice that the interaction modifies the mean particle number since the term $u^{\prime}\left( \bar{n}\right)$ is present in Eq. (\ref{eq:Mean-number-N-new-n1}). This modification is directly related to the fact that we chose the interaction energy to be a function of the particle density.

Next, the entropy is given by
\begin{align}
S& =-\left. \frac{\partial \Phi }{\partial T}\right\vert _{\mu ,V},%
\displaybreak[0]  \notag \\
& =-V\left. \frac{\partial u\left( \bar{n}\right) }{\partial T}\right\vert
_{\mu ,V}+\bar{N}\left. \frac{\partial u^{\prime }\left( \bar{n}\right) }{%
\partial T}\right\vert _{\mu ,V}+u^{\prime }\left( \bar{n}\right) \left.
\frac{\partial \bar{N}}{\partial T}\right\vert _{\mu ,V}\displaybreak[0]
\notag \\
& +\chi \sum_{\Omega }\ln \left( 1+\chi \exp \left[ -\beta \left( E_{\Omega
}+u^{\prime }\left( \bar{n}\right) -\mu \right) \right] \right) %
\displaybreak[0]  \notag \\
& +\chi ^{2}T\sum_{\Omega }\frac{\left( E_{\Omega }+u^{\prime }\left( \bar{n}%
\right) -\mu \right) \left( -\frac{d\beta }{dT}\right) -\beta \left. \frac{%
\partial u^{\prime }\left( \bar{n}\right) }{\partial T}\right\vert _{\mu ,V}%
}{\exp \left[ \beta \left( E_{\Omega }+u^{\prime }\left( \bar{n}\right) -\mu
\right) \right] +\chi },
\end{align}%
or in a more compact form,
\begin{eqnarray}
S &=&\chi \sum_{\Omega }\ln \left( 1+\chi \exp \left[ -\beta \left(
E_{\Omega }+u^{\prime }\left( \bar{n}\right) -\mu \right) \right] \right)
\notag \\
&&+\frac{1}{T}\sum_{\Omega }\bar{n}_{\Omega }\left( E_{\Omega }+u^{\prime
}\left( \bar{n}\right) -\mu \right) .  \label{eq:Entropy1}
\end{eqnarray}
Moreover, the mean energy reads%
\begin{align}
\bar{E}& =\left. \frac{\partial \left( \beta \Phi \right) }{\partial \beta }%
\right\vert _{z,V},\displaybreak[0]  \notag \\
& =\left. \frac{\partial }{\partial \beta }\left[ \beta Vu\left( \bar{n}%
\right) -\beta \bar{N}u^{\prime }\left( \bar{n}\right) \right] \right\vert
_{z,V}\displaybreak[0]  \notag \\
& -\chi \sum_{\Omega }\frac{\chi z\exp \left[ -\beta \left( E_{\Omega
}+u^{\prime }\left( \bar{n}\right) \right) \right] }{1+\chi z\exp \left[
-\beta \left( E_{\Omega }+u^{\prime }\left( \bar{n}\right) \right) \right] }%
\left( -E_{\Omega }-\left. \frac{\partial }{\partial \beta }\left[ \beta
u^{\prime }\left( \bar{n}\right) \right] \right\vert _{z,V}\right) ,%
\displaybreak[0]  \notag \\
& =U\left( V,\bar{n}\right) +\beta V\left. \frac{\partial u\left( \bar{n}%
\right) }{\partial \beta }\right\vert _{z,V}-\bar{N}\left. \frac{\partial %
\left[ \beta u^{\prime }\left( \bar{n}\right) \right] }{\partial \beta }%
\right\vert _{z,V}-\beta u^{\prime }\left( \bar{n}\right) \left. \frac{%
\partial \bar{N}}{\partial \beta }\right\vert _{z,V}\displaybreak[0]  \notag
\\
& +\sum_{\Omega }\frac{E_{\Omega }}{z^{-1}\exp \left[ \beta \left( E_{\Omega
}+u^{\prime }\left( \bar{n}\right) \right) \right] +\chi }+\bar{N}\left.
\frac{\partial \left[ \beta u^{\prime }\left( \bar{n}\right) \right] }{%
\partial \beta }\right\vert _{z,V}.  \label{eq:mean-energy}
\end{align}%
After some algebraic manipulations, one can rewrite this expression as%
\begin{equation}
\bar{E}=\sum_{\Omega }\bar{n}_{\Omega }E_{\Omega }+U\left( V,\bar{n}\right) .
\label{eq:Energy1}
\end{equation}
This is an expected result since the energy is the average of the kinetic term plus the interactions energy. Finally, we derive the pressure of the system%
\begin{align}
p& =-\left. \frac{\partial \Phi }{\partial V}\right\vert _{\mu ,T},%
\displaybreak[0]  \notag \\
& =-u\left( \bar{n}\right) +u^{\prime }\left( \bar{n}\right) \left. \frac{%
\partial \bar{N}}{\partial V}\right\vert _{\mu ,T}+\frac{\chi T}{V}%
\sum_{\Omega }\ln \left( 1+\chi \exp \left[ -\beta \left( E_{\Omega
}+u^{\prime }\left( \bar{n}\right) -\mu \right) \right] \right) ,%
\displaybreak[0]  \label{eq:pressurePP} \\
& =-\frac{\Phi }{V},
\end{align}%
where we have used Eq. (\ref{eq:Mean-number-N}) and the fact that the particle density does not depend on the volume. From Eq. (\ref{eq:pressurePP}), we can also realize how interaction plays an important role on the pressure. The first term in Eq. (\ref{eq:pressurePP}), - $u\left( \bar{n}\right)$, for instance, is responsible to reduce the pressure of the system, while the second one, $u^{\prime }\left( \bar{n}\right)$, plays the role of increasing it instead. It is worth mentioning that such thermal functions were recently calculated in \cite{tt0,tt1,tt2,tt3,tt4,tt6,aa2022particles} for Lorentz-violating systems.

\subsection{Analytical results for three-dimensional boxes}\label{Sec:Ex1}

We exemplify our model constructed above for the three-dimensional box. The spectral energy is%
\begin{equation}
E_{\eta _{x},\eta _{y},\eta _{z}}^{\text{\textrm{Box}}}=\frac{\pi ^{2}\hbar
^{2}}{2\mathrm{M}}\left( \frac{\eta _{x}^{2}}{L_{x}^{2}}+\frac{\eta _{y}^{2}}{%
L_{y}^{2}}+\frac{\eta _{z}^{2}}{L_{z}^{2}}\right) .
\end{equation}%
Here, the grand canonical potential of an \textit{interacting} gas is%
\begin{equation}
\Phi =-k_{B}T\chi \sum_{\left\{ \eta _{x},\eta _{y},\eta _{z}\right\} }\ln
\left\{ 1+\chi \exp \left[ -\beta \left( E_{\eta _{x},\eta _{y},\eta _{z}}^{%
\text{\textrm{Box}}}+u^{\prime }\left( \bar{n}\right) -\mu \right) \right]
\right\} +U\left( V,\bar{n}\right) -u^{\prime }\left( \bar{n}\right) \bar{N}.
\label{eq:newnewphi}
\end{equation}%
To proceed further, the \textit{Euler--MacLaurin formula} \cite{ada1,ada2,tt5} must be utilized,%
\begin{align}
\sum_{n}^{\infty }F\left( n\right) & =\int_{0}^{\infty }F\left( n\right) dn+%
\frac{1}{2}F\left( 0\right) \displaybreak[0]  \notag \\
& -\frac{1}{2!}B_{2}F^{\prime }\left( 0\right) -\frac{1}{4!}B_{4}F^{\prime
\prime \prime }\left( 0\right) +\ldots ,
\end{align}
where $B_{k}'s$ are the Bernoulli numbers. This allows us to approximate $\Phi$, for low-$T$ regime\footnote{ The higher order terms become negligible because the low temperature regime under consideration makes the exponential function very small in Eq. (\ref{eq:newnewphi}). The higher order corrections become relevant for high-$T$ regime. On the other hand, for low-$T$ regime, we can estimate $T<100~\mathrm{K}$.}, as:%
\begin{align}
\Phi & \approx -k_{B}T\chi \int_{0}^{\infty }\int_{0}^{\infty }\int_{0}^{\infty }d\eta
_{x}d\eta _{y}d\eta _{z}\ln \left\{ 1+\chi \mathfrak{z}\exp \left[ -\beta
E_{\eta _{x},\eta _{y},\eta _{z}}^{\text{\textrm{Box}}}\right] \right\} %
\displaybreak[0]  \notag \\
& +\frac{k_{B}T\chi }{2}\int_{0}^{\infty }\int_{0}^{\infty }d\eta _{y}d\eta
_{z}\ln \left\{ 1+\chi \mathfrak{z}\exp \left[ -\beta E_{0,\eta _{y},\eta
_{z}}^{\text{\textrm{Box}}}\right] \right\} \displaybreak[0]  \notag \\
& +\frac{k_{B}T\chi }{2}\int_{0}^{\infty }\int_{0}^{\infty }d\eta _{x}d\eta
_{z}\ln \left\{ 1+\chi \mathfrak{z}\exp \left[ -\beta E_{\eta _{x},0,\eta
_{z}}^{\text{\textrm{Box}}}\right] \right\} \displaybreak[0]  \notag \\
& +\frac{k_{B}T\chi }{2}\int_{0}^{\infty }\int_{0}^{\infty }d\eta _{x}d\eta
_{y}\ln \left\{ 1+\chi \mathfrak{z}\exp \left[ -\beta E_{\eta _{x},\eta
_{y},0}^{\text{\textrm{Box}}}\right] \right\} \displaybreak[0]  \notag \\
& +\frac{k_{B}T\chi }{4}\int_{0}^{\infty }d\eta _{x}\ln \left\{ 1+\chi \mathfrak{z%
}\exp \left[ -\beta E_{\eta _{x},0,0}^{\text{\textrm{Box}}}\right] \right\} %
\displaybreak[0]  \notag \\
& +\frac{k_{B}T\chi }{4}\int_{0}^{\infty }d\eta _{y}\ln \left\{ 1+\chi \mathfrak{z%
}\exp \left[ -\beta E_{0,\eta _{y},0}^{\text{\textrm{Box}}}\right] \right\} %
\displaybreak[0]  \notag \\
& +\frac{k_{B}T\chi }{4}\int_{0}^{\infty }d\eta _{z}\ln \left\{ 1+\chi \mathfrak{z%
}\exp \left[ -\beta E_{0,0,\eta _{z}}^{\text{\textrm{Box}}}\right] \right\} %
\displaybreak[0]  \notag \\
& +\frac{k_{B}T\chi }{8}\ln \left( 1+\chi \mathfrak{z}\right) +U\left( V,\bar{n}%
\right) -u^{\prime }\left( \bar{n}\right) \bar{N},
\end{align}%
where we have defined $\mathfrak{z}=ze^{-\beta u^{\prime }\left( \bar{n}%
\right) }$. After performing the integrals, we obtain%
\begin{equation}
\Phi = \frac{\mathcal{V}}{\lambda ^{3}}h_{\frac{5}{2}}\left( \mathfrak{z}%
\right) -\frac{1}{4}\frac{\mathcal{S}}{\lambda ^{2}}h_{2}\left( \mathfrak{z}%
\right) +\frac{1}{16}\frac{\mathcal{L}}{\lambda }h_{\frac{3}{2}}\left(
\mathfrak{z}\right) -\frac{1}{8}h_{1}\left( \mathfrak{z}\right) +U\left( V,%
\bar{n}\right) -u^{\prime }\left( \bar{n}\right) \bar{N},  \label{eq:3Dcase}
\end{equation}%
where $\lambda =h/\sqrt{2\pi \mathrm{M} k_{B}T}$ is the thermal wavelength, $\mathcal{V%
}=L_{x}L_{y}L_{z}$ is the volume, $\mathcal{S}=2\left(
L_{x}L_{y}+L_{y}L_{z}+L_{z}L_{x}\right) $ the area of the surface, $%
\mathcal{L}=4\left( L_{x}+L_{y}+L_{z}\right) $ the total length of the edges of the box and
\begin{equation}
h_{\sigma }\left( \mathfrak{z}\right) =\frac{1}{\Gamma \left( \sigma \right)
}\int_{0}^{\infty }\frac{t^{\sigma -1}}{\mathfrak{z}^{-1}e^{t}+\chi }%
dt=\left\{
\begin{array}{l}
f_{\sigma }\left( \mathfrak{z}\right) ,\text{ for fermions }\left( \chi
=1\right)  \\
g_{\sigma }\left( \mathfrak{z}\right) ,\text{ for bosons }\left( \chi
=-1\right)
\end{array}%
\right. .
\end{equation}%
The boundary effects in Eq. (\ref{eq:3Dcase}) are represented by the second and third terms which are proportional to the perimeter $\mathcal{L}$ and to the
surface $\mathcal{S}$. We note that these terms are modified by the interaction term $u^{\prime }\left( \bar{n}\right)$. Also, we can carry out a similar calculation involving a two-dimensional box. 

As an application, let us use the result obtained from Eq. (\ref{eq:3Dcase}) to probe how interaction affects the Fermi energy. From that, we get

\begin{equation}
N=g\left[ \frac{V}{\lambda ^{3}}f_{\frac{3}{2}}\left( \mathfrak{z}\right) -%
\frac{1}{4}\frac{S}{\lambda ^{2}}f_{1}\left( \mathfrak{z}\right) +\frac{L}{%
16\lambda }f_{\frac{1}{2}}\left( \mathfrak{z}\right) -\frac{1}{8}\frac{%
\mathfrak{z}}{\mathfrak{z}+1}\right] ,
\end{equation}%
where g is a weight factor that arises from the internal structure of the
particles. The Fermi energy $\mu _{0}$ is the energy of the topmost filled
level in the ground state of an electron system. In this way,%
\begin{equation}
N=g\left[ \frac{V}{6\pi ^{2}}\left( \frac{2m\mathfrak{X}}{\hbar ^{2}}\right) ^{%
\frac{3}{2}}-\frac{1}{4}%
\frac{S}{4\pi }\left( \frac{2m\mathfrak{X}}{\hbar ^{2}}\right)+\frac{L}{16\pi }\left( \frac{2m\mathfrak{X}}{\hbar
^{2}}\right) ^{\frac{1}{2}}-%
\frac{1}{8}\right] ,
\label{eq:Nresults3}
\end{equation}%
where $\mathfrak{X}=\mu _{0}-u^{\prime}\left( n\right)$. The Fermi energy cannot be calculated from this equation until we conveniently choose the interaction $u(n)$ term. However, we can at least see how the interaction modifies the
structure of the equation which determines the Fermi energy $\mu _{0}$. It is
also possible to infer that the existence of a interaction enhances such energy level. Moreover, we realize that the interaction remarkably introduces a
density-dependence on the Fermi energy. Unfortunately, even for a linear aproximation, it is not possible to get a analytical result for $\mu_{0}$. On the other hand, Eq. (\ref{eq:Nresults3}) give us an idea of the challenge that we have to face in order to solve $\mu_{0}$ numerically.

Finally, straightforward question is worthy to be taken into account: how would the thermal properties be influenced if one considered spherical (\ref{sphe}), cylindrical (\ref{cyli}), and ellipsoidal (\ref{ellip}) potentials instead? Despite being an intriguing question, it is challenging to perform such an analysis by analytical means. Nevertheless, this analysis will be performed numerically in an upcoming work.


\section{Conclusion and future perspectives}\label{conclusion}

We examined the behavior of the thermodynamic functions for different geometries, i.e., spherical, cyllindrical, and ellipsoidal ones; we primarily used the canonical ensemble for spinless particles. Moreover, \textit{noninteracting} gases were also taken into account for the same geometries with the usage of the grand canonical ensemble description. A study of how geometry affected the system of spinless particles, fermions and bosons was provided as well. We also note that our results could possibly be applied to the \textit{Bose-Einstein condensate} and to the \textit{helium dimer}.

Furthermore, for the bosonic sector, independently of the geometry, the entropy and internal energy turned out to be greater than for the fermionic case; a standard ordering of the sizes of the computed quantities repeatedly occurred for both systems: $\textit{Ellipsoid}>\textit{Cylinder}>\textit{Sphere}$. 

Finally, we constructed a model to provide a description of \textit{interacting} quantum gases;  it was implemented for a cubical box. Such an interaction sector turned out to be more prominent since the results were derived analytically. More so, it was possible to estimate the Fermi energy for the cubical box and see how the interaction played the role of modifying it. We also shown how to build an approximated interaction term in order to estimate such energy. Nevertheless, even for the simplest case, we could not obtain analytical results. This aspect clearly demonstrated how intricate were the calculations when we took the interaction into account. Furthermore, another remarkable feature worth exploring, would be the thermodynamic aspects of anisotropic systems \cite{kostelecky}.


\section*{Acknowledgments}
\hspace{0.5cm}

The authors would like to thank João Milton, Andrey Chaves, Diego Rabelo, Ewerton Wagner and Paulo Porfírio for the fruitful suggestions during the preparation of this manuscript. More so, we are also indebted to Albert Petrov, Izeldin Ahmed, Marco Schreck, Subir Ghosh, anonymous referee, and the editor for the corrections and recommendations given to this work. Particularly, A. A. Araújo Filho acknowledges the Facultad de Física - Universitat de València and Gonzalo J. Olmo for the kind hospitality when part of this work was made. Moreover, this work was partially supported by Conselho Nacional de Desenvolvimento Cient\'{\i}fico e Tecnol\'{o}gico (CNPq) - 142412/2018-0, Coordenação de Aperfeiçoamento de Pessoal de Nível Superior (CAPES) - Finance Code 001, and CAPES-PRINT (PRINT - PROGRAMA INSTITUCIONAL DE INTERNACIONALIZAÇÃO) - 88887.508184/2020-00. 




\bibliographystyle{ieeetr}
\bibliography{main}

\end{document}